\documentclass[11pt,notitlepage]{article}
\usepackage{amssymb}
\usepackage{psfig}
\usepackage[dvips]{graphics}
\usepackage{subeqnarray}
\usepackage{graphicx}
\usepackage{subeqnarray}
\usepackage{subfigure}

\newcommand{\be}{\begin{equation}}
\newcommand{\ee}{\end{equation}}
\newcommand{\bea}{\begin{eqnarray}}
\newcommand{\eea}{\end{eqnarray}}

\title{\bf Degenerate Four Virtual Soliton Resonance for KP-II}

\author{Oktay K. Pashaev and
Meltem L. Y. Francisco \\ \\
Department of Mathematics,
Izmir Institute of Technology \\
Urla-Izmir, 35430, Turkey}

\begin{document}
\maketitle

\begin{abstract}

 By using disipative version of the second and
the third members of AKNS hierarchy, a new method to solve 2+1
dimensional Kadomtsev-Petviashvili (KP-II) equation is proposed.
We show that dissipative solitons (dissipatons) of those members
give rise to the real solitons of KP-II. From the Hirota bilinear
form of the SL(2,R) AKNS flows, we formulate a new bilinear
representation for KP-II, by which, one and two soliton solutions
are constructed and the resonance character of their mutual
interactions is studied. By our bilinear form, we first time
created four virtual soliton resonance solution for KP-II and
established relations  of it with degenerate four-soliton
solution in the Hirota-Satsuma bilinear form for KP-II.
\end{abstract}

~
\newtheorem{thm}{Theorem}[subsection]
\newtheorem{cor}[thm]{Corollary}
\newtheorem{lem}[thm]{Lemma}
\newtheorem{prop}[thm]{Proposition}
\newtheorem{defn}[thm]{Definition}
\newtheorem{rem}[thm]{Remark}

\section{Introduction}

Recently, dissipative version of AKNS hierarchy has been
considered in connection with 1+1 dimensional (lineal) gravity
models \cite{MPS}. It is found that the second flow described by
the dissipative version of the Nonlinear Schr\"odinger equation -
the so called reaction-diffusion system, admits new soliton type
solutions called {\it dissipatons}. Dissipatons have exponentially
growing-decaying amplitudes, with perfect soliton shape for the
bilinear product of them and the resonance interaction behaviour.

In the present paper we study resonance dissipative solitons in
AKNS hierarchy and show that they give rise to the real solitons
of KP-II.  Our approach is based on a new method to generate
solutions of 2+1 dimensional  KP equation: we show that if one
considers a simultaneous solution of the second and the third
flows from the AKNS hierarchy, then the product $e^+ e^-$
satisfies the KPII equation (Proposition 1). Using these results
we construct new bilinear representation of KPII equation with
one and two soliton solutions. We show that our two-soliton
solution corresponds to the degenerate four soliton solution in
the standard Hirota form of KP , and displays the four virtual
soliton  resonance.

\section{SL(2,R) AKNS Hierarchy}

The dissipative SL(2,R) AKNS hierarchy of evolution equations
\be \frac{1}{2}\sigma_3  \left (\begin{array}{clcr}e^+ \\
e^- \end{array} \right)_{t_{N}}= \Re ^{N+1} \left (\begin{array}{clcr} e^+\\
e^- \end{array} \right) ,\label{hier}\ee where $N = 0,1,2,...$,
($\Lambda < 0$), is generated by the recursion operator $ \Re $
\be \Re = \left(\begin{array}{cccr}\partial_x-\frac{\Lambda}{4}e^+
\int^x e^- & -\frac{\Lambda}{4}e^+ \int^x e^+
\\ & \\-\frac{\Lambda}{4}e^- \int^x
e^-&\partial_x+\frac{\Lambda}{4}e^- \int^x e^+\end{array}
\right).\ee Then, the second and  third members of AKNS hierarchy
appear as \be\left\{\begin{array}{c}
e^+_{t_{1}} = e^+_{xx} +\frac{\Lambda}{4}e^+ e^- e^+ \\
-e^-_{t_{1}} = e^-_{xx} +\frac{\Lambda}{4}e^+ e^-
e^-\end{array}\label{pair1}\right.\ee and \be\left\{
\begin{array}{c}
e^+_{t_{2}} = e^+_{xxx} +\frac{3\Lambda}{4}e^+ e^-  e^+_{x} \\
e^-_{t_{2}} = e^-_{xxx} +\frac{3\Lambda}{4}e^+e^- e^-_{x}
\end{array}\label{pair2}\right. \ee
respectively. The first system (\ref{pair1}), the dissipative
version of the Nonlinear Schr\"odinger equation, is called the
Reaction-Diffusion (RD) system \cite{MPS}. It is connected with
gauge theoretical formulation of 1+1 dimensional gravity, the
constant curvature surfaces in pseudo-Euclidean space \cite{MPS}
and the NLS soliton problem in the quantum potential \cite{MPS}
\cite{resonance}.

\section{Resonance Dissipatons in AKNS Hierarchy}

\subsection{Dissipatons of Reaction-Diffusion System}

The second member of the AKNS (\ref{pair1}), the
Reaction-Diffusion equation, by substitution \be
e^{\pm}=\sqrt{\frac{8}{-\Lambda}}\frac{G^\pm(x,t)}{F(x,t)},\label{sub}\ee
admits the Hirota bilinear representation, $(t \equiv t_1)$, \be
(\pm D_t-D_{x}^{2})(G^\pm\cdot F)=0,\,\,\,\, D_{x}^{2}(F\cdot
F)=-2G^+G^-. \label{bi-RD1} \ee Then, any solution of this system
determines a solution of the Reaction-Diffusion system
(\ref{pair1}). Simplest solution of bilinear system
(\ref{bi-RD1}),  has  the form \cite{resonance} \be G^{\pm} = \pm
e^{ \eta_{1}^{\pm}},\hskip 0.5cm F = 1 + \frac {e^{(\eta_{1}^{+}
+ \eta_{1}^{-})}}{(k_{1}^{+}+ k_{1}^{-})^2}, \label{1dis}\ee where
$\eta _{1}^{\pm } = k_{1}^{\pm}x \pm  (k_{1}^{\pm})^2 t + \eta
^{\pm(0)}_{1}$. This solution determines soliton-type solution of
the Reaction-Diffusion system  with exponentially growing and
decaying amplitudes, called the dissipaton \cite{MPS}. But for
the product $e^+ e^-$ one have the perfect one-soliton shape \be
e^+ e^- = \frac{8 k^{2}}{\Lambda \cosh^2[k(x - v t - x_0)]},\ee of
the amplitude $k = (k_{1}^{+} + k_{1}^{-})/2$, propagating with
velocity $v = -(k_1^+ - k_{1}^{-})$, where the initial position $
x_0 = -\ln(k_{1}^{+} + k_{1}^{-})^2 +
\eta_{1}^{+(0)}+\eta_{1}^{-(0)} $.

The Reaction-Diffusion system has a geometrical interpretation in
a language of constant curvature surfaces \cite{MPS}.  It follows
that when $e^\pm$ satisfy  the Reaction-Diffusion equations
(\ref{pair1}), the Rimannian  metric describes two-dimensional
pseudo-Riemannian space-time with constant curvature $\Lambda$: $R
= g^{\mu\nu}R_{\mu\nu} = \Lambda$. If we calculate the metric for
one dissipaton solution (\ref{1dis}) it shows a singularity (sign
changing) at $\tanh k(x-vt)= \pm v/2k$. This singularity (called
the causal singularity )has physical interpretation in terms of
black hole physics and relates with resonance properties of
solitons. In fact, constructing two dissipaton solution we find
that it describes a collision of two dissipatons creating the
resonance (metastable) bound state \cite{resonance}.

\subsection{Dissipative Solitons for the Third Flow}

For the third flow of AKNS hierarchy we have the qubic dispersion
system (\ref{pair2}). The bilinear representation of this system
for functions $e^{\pm}(x,t)$  in terms of three real functions
$G^\pm, F$, as in (\ref{sub}), is \be ( D_t+D_{x}^{3})(G^\pm\cdot
F)=0,\,\,\,\,D_{x}^{2}(F\cdot F)= -2G^+G^-. \label{hir31}\ee From
the last equation we have expression for the product \be
U=e^{+}e^{-}=\frac{8}{-\Lambda}\frac{G^+G^-}{F^2}=\frac{4}{\Lambda}\frac{D_{x}^{2}(F\cdot
F)}{F^2}= \frac{8}{\Lambda}\frac{\partial^2}{\partial x^2}\ln F
.\ee Simplest solution of this system  \be G^{\pm} = \pm e^{
\eta_{1}^{\pm}}, F = 1 + \frac {e^{(\eta_{1}^{+} +
\eta_{1}^{-})}}{(k_{1}^{+}+ k_{1}^{-})^2} ,\ee where  $\eta
_{1}^{\pm } = k_{1}^{\pm}x  - (k_{1}^{\pm})^3t + \eta
^{\pm(0)}_{1}$, defines one dissipative soliton solution of the
system (\ref{pair2}) \be e^\pm =
\pm\sqrt{\frac{8}{-\Lambda}}\frac{|k_{11}^{\pm}|}{2}\frac{e^{\pm
\frac{1}{2}(\eta_1^+ - \eta_1^-)}}{\cosh\frac{k_1^+ + k_1^-
}{2}[x-v t -x_{0}]},\label{dis3}\ee where $ v =
{k_1^+}^{2}-k_1^+k_1^- +{k_1^-}^{2}$, $x_0 = ({\eta_1^+}^{(0)}+
{\eta_1^-}^{(0)})/k_1^+ k_1^-$,$\phi_{11} = -2 \ln k^{+-}_{11} $.

The system (\ref{pair2}) admits following symmetric reduction: $
e^+ = e^- = u$, leading to the MKdV equation \be u_{t_2}= u_{xxx}
+ \frac{3\Lambda}{4}u^2 u_x. \label{mkdv}\ee Under this
reduction, $k^+_1 = k^-_1 \equiv k$ , and the dissipative soliton
(\ref{dis3}) becomes one-soliton solution of MKdV \be e^+ = e^- =
u(x,t) = \sqrt{\frac{8}{-\Lambda}}\frac{|k|}{\cosh k(x-k^2 t -
x_0)}.\ee This way we can see that dissipative soliton is a more
general object reducible to the real soliton of MKdV. In a
similar way two dissipative soliton solution of system
(\ref{pair2}) under reduction $k^+_1 = k^-_1$, $k^+_2 = k^-_2$,
is reducible to two soliton solution of MKdV. The natural
question is to find an evolution equation for $e^+ e^-$ product
of dissipatons. As we show below it is KPII equation in 2+1
dimensions.

\section{KP-II Resonance Solitons}

\subsection{KP-II and AKNS hierarchy}

AKNS hierarchy allows us to develop also a new method to find
solution for (2+1)  Kadomtsev-Petviashvili (KP)equation.
Depending on sign of dispersion, two types of the KP equations
are known. The minus sign in the right side of the KP corresponds
to the case of negative dispersion and called KPII. To relate
KPII with AKNS hierarchy let us consider the pair of functions
$e^+(x,y,t)$, $e^-(x,y,t)$ satisfying the second
 and the third
members of the dissipative AKNS hierarchy. Here we renamed  time
variables $t_1$ as $y $ and $t_2$ as $t$. Differentiating
according to t and y, Eqs. (\ref{pair1}) and (\ref{pair2})
correspondingly,  we can see that they are compatible.
\begin{prop}
Let the functions $e^+(x,y,t)$ and $e^-(x,y,t)$ are solutions of
equations (\ref{pair1}) and (\ref{pair2}) simultaneously. Then
the function $U(x,y,t) \equiv e^+e^-$ satisfies the
Kadomtsev-Petviashivili (KPII) equation
\be(4U_t+\frac{3\Lambda}{4}(U^2)_x+U_{xxx})_x =
-3U_{yy}.\label{kp2}\ee

\end{prop}
\noindent {\bf Proof:} We take derivative of $U$ according to $y$
variable and use Eq.(\ref{pair1}), so that $ U_y =  {(e^+_{x}e^- -
e^-_{x}e^+ )_x}$,  \be U_{yy} = {(e^+_{xxx}e^- + e^-_{xxx}e^+ -
(e^+_xe^-_x)_x)}+\frac{\Lambda}{2}U_xU . \ee In a similar way for
$U_t$ we have \be U_t = -(e^+_{xxx}e^- +
\frac{3\Lambda}{4}Ue^-e^-_x + e^-_{xxx}e^+ + \frac{3\Lambda}{4}U
e_x^-e^+), \ee  \be U_{xt}= - (e_{xxx}^+e^- + e_{xxx}^-e^+ +
\frac{3\Lambda}{4}UU_x )_x. \ee Combining above formulas together
\be 4U_{xt} + 3U_{yy} = [-e_{xxx}^+e^- - e_{xxx}^-e^+ -
\frac{3\Lambda}{2}UU_x - 3(e^+_xe^-_x)_x]_x,\ee and using $U_{xxx}
= e_{xxx}^+e^- + e_{xxx}^-e^+ + 3e^+_{xx}e^-_x + 3e^+_xe^-_{xx}$,
we get KPII (\ref{kp2})\footnote[1]{As Konopelchenko recently
mentioned to us the similar results are known also as the symmetry
reductions of KP \cite{KS},\cite{ChengLi},\cite{cao} }.

\subsection{Bilinear Representation of KPII by AKNS flows}

Using bilinear representations for systems (\ref{pair1}) and
(\ref{pair2}) and Proposition $\bf 4.1.1$ we can find bilinear
representation for KPII. For RD system (\ref{pair1}) bilinear
form is given by (\ref{bi-RD1}), while for the third flow
(\ref{pair2}) by Eqs. (\ref{hir31}).

 Now we consider $G^{\pm}$ and $F$ as functions of
three variables $G^{(\pm)} = G^{(\pm)}(x,y,t)$, $F = F(x,y,t)$,
and require for these functions  to be a solution of both bilinear
systems (\ref{bi-RD1}), (\ref{hir31}) simultaneously. Since the
second equation in both systems is the same, it is sufficient to
consider the next bilinear system \be \left\{
\begin{array}{llcr}(\pm D_y-D_{x}^{2})(G^\pm\cdot F)=0 \\(
D_t+D_{x}^{3})(G^\pm\cdot F)=0
\\D_{x}^{2}(F\cdot F)=-2G^+G^-
\end{array} \label{hirform}\right.\ee
Then, according to Proposition $\bf {4.1.1}$, any solution of this
system generates a solution of KPII. From the last equation we
can derive $U$ directly in terms of function $F$ only \be
U=e^{+}e^{-}=\frac{8}{-\Lambda}\frac{G^+G^-}{F^2}=\frac{4}{\Lambda}\frac{D_{x}^{2}(F\cdot
F)}{F^2}= \frac{8}{\Lambda}\frac{\partial^2}{\partial x^2}\ln F
\label{Uln}\ee Simplest solution of this system  \be G^{\pm} =
\pm e^{ \eta_{1}^{\pm}},\,\, F = 1 + \frac {e^{(\eta_{1}^{+} +
\eta_{1}^{-})}}{(k_{1}^{+}+ k_{1}^{-})^2} ,\ee where  $\eta
_{1}^{\pm } = k_{1}^{\pm}x \pm (k_{1}^{\pm})^2y -
(k_{1}^{\pm})^3t + \eta ^{\pm(0)}_{1}$, defines one-soliton
solution of KPII according to Eq.(\ref{Uln}) \be U =
\frac{2(k_{1}^{+} + k_{1}^{-})^{2}}{\Lambda \cosh^2
\frac{1}{2}[(k_{1}^{+} + k_{1}^{-})x + ({k_{1}^{+}}^{2} -
{k_{1}^{-}}^{2})y -({k_{1}^{+}}^{3} +
{k_{1}^{-}}^{3})t+\gamma]},\label{1ss}\ee where $ \gamma =
-\ln(k_{1}^{+} + k_{1}^{-})^2 + \eta_{1}^{+(0)}+\eta_{1}^{-(0)} $.
This soliton is a planar wave wall traveling in an arbitrary
direction and called the line soliton.

\subsection{Two Soliton Solution}

Continuing Hirota's perturbation we find two soliton solution in
the form

\be G^{\pm} = \pm (e ^{\eta_{1}^{\pm}} + e^{ \eta_{2}^{\pm}}+
\alpha_{1} ^{\pm}e ^{{\eta_{1}^{+}+ \eta_{1}^{-}+\eta_{2}^{\pm}}}+
\alpha_{2} ^{\pm}e^{ {\eta_{2}^{+}+ \eta_{2}^{-} +
\eta_{1}^{\pm}}} ),\label{2ssa}\ee

\be F = 1 + \frac{e^{ \eta_{1}^{+} +
\eta_{1}^{-}}}{(k_{11}^{+-})^2} + \frac{e^{ \eta_{1}^{+}
+\eta_{2}^{-}}}{(k_{12}^{+-})^2} + \frac{e^{ \eta_{2}^{+} +
\eta_{1}^{-}}}{(k_{21}^{+-})^2} + \frac{e^{ \eta_{2}^{+}+
\eta_{2}^{-}}}{(k_{22}^{+-})^2 }+ \beta e^{{\eta_{1}^{+} +
\eta_{1}^{-} + \eta_{2}^{+} + \eta_{2}^{-}} }, \label{2ssb}\ee
where  $ \eta _{i}^{\pm } = k_{i}^{\pm}x  \pm (k_{i}^{\pm})^2y -
(k_{i}^{\pm})^3 t + \eta ^{\pm(0)}_{i}$, $k_{ij}^{ab} = k_{i}^{a}
+ k_{j}^{b} $, $(i,j= 1,2)$, $ (a,b = + -)$,
$$\alpha_{1}^{\pm} = \frac{(k_{1}^{\pm} -
k_{2}^{\pm})^{2}}{(k_{11}^{+-}k_{21}^{\pm\mp})^2}, \hskip 0.5cm
\alpha_{2}^{\pm}= \frac{(k_{1}^{\pm}-
k_{2}^{\pm})^2}{(k_{22}^{+-}k_{12}^{\pm \mp})^2}, \hskip0.5cm\beta
= \frac{(k_{1}^{+}- k_{2}^{+})^2(k_{1}^{-}-
k_{2}^{-})^2}{(k_{11}^{+-}k_{12}^{+-}k_{21}^{+-}k_{22}^{+-})^2}.$$
 It provides two-soliton solution of
KPII according to Eq.(\ref{Uln}).

\subsection{Degenerate Four-Soliton Solution}

For KPII another bilinear form in terms of function F only is
known \cite{hirota1} \be( D_xD_t + D_{x}^{4} + D_{y}^{2})(F\cdot
F) = 0 \label{biwadati}\ee Thus, it is natural to compare soliton
solutions of our bilinear equations (\ref{hirform}) with the ones
given by this equation. To solve equation (\ref{biwadati}) we
consider $ F = 1 + \varepsilon F_{1} + \varepsilon^2 F_{2} + ...$.
The solution $ F_1 = e^{\eta_1}$, where $\eta _{1} = k_{1}x +
\Omega_{1}y + \omega_{1} t + \eta _{1}^{0}$, and dispersion $ k_1
\omega_1 + k_{1}^{4}+\Omega_{1}^{2} = 0$ with $F_n = 0$, (n =
2,3,...), under identification $k_1 = k_{1}^{+} + k_{1}^{-}$,
$\Omega_1 = \sqrt{3}(k_{1}^{+2} - k_{1}^{-2} )$, $\omega_1 =
-4(k_{1}^{+3} + k_{1}^{-3} )$, and rescaling $4 t \rightarrow t$,
$\sqrt{3} y \rightarrow y$, determines one soliton solution of
KPII (\ref{kp2}). We realize that it coincides with our one
soliton solution (\ref{1ss}).  But two soliton solution of
equation (\ref{biwadati}) is not correspond to our two-soliton
solution (\ref{2ssa}),(\ref{2ssb}). Appearance of four different
terms $e^{\eta^{\pm}_{i} + \eta^{\pm}_{k}}$  in equation
(\ref{2ssb}), suggest  that our two-soliton solution should
corresponds to some degenerate case of four soliton solution of
Eq(\ref{biwadati}).\footnote[2]{One of the authors (O.P.) thanks
Prof. J. Hietarinta for this suggestion} To construct four
soliton solution first we find following solutions of bilinear
equations (\ref{biwadati})

 \be F_1 = e^{\eta_1},\,\, F_2 = e^{\eta_2},\,\,F_4 = e^{\eta_3},\,\,
 \ee where
$ \eta _{i} = k_{i}x + \Omega_{i}y + \omega_{i} t + \eta
_{i}^{0}$,$ i = 1,2,3,$ dispersion \be k_i \omega_i +
k_{i}^{4}+\Omega_{i}^{2} = 0 \label{disp}\ee
 and
\be F_3 = \alpha_{12}e^{\eta_1 + \eta_2},\,\,F_5 =
\alpha_{13}e^{\eta_1 + \eta_3},\,\, F_6 = \alpha_{23}e^{\eta_1 +
\eta_3},\,\,
 \ee where \be \alpha_{ij} =
-\frac{(k_i - k_j)(\omega_i-\omega_j)+ {(k_i-k_j)}^4 +
{(\Omega_i-\Omega_j)}^2}{(k_i + k_j)(\omega_i+\omega_j)+
{(k_i+k_j)}^4 + {(\Omega_i+\Omega_j)}^2}, i,j = 1,2,3.\ee Then we
parameterize  our solution in the form \be
\begin{array} {cccc} k_1 = k_{1}^{+} + k_{1}^{-}, & \omega_1 = -4(k_{1}^{+3} + k_{1}^{-3}
), & \Omega_1 = \sqrt{3}(k_{1}^{+2} - k_{1}^{-2} ),
\\k_2 =
k_{2}^{+} + k_{2}^{-},&  \omega_2 = -4(k_{2}^{+3} + k_{2}^{-3} ),
& \Omega_2 =
\sqrt{3}(k_{2}^{+2} - k_{2}^{-2} ),
\\ k_3 = k_{1}^{+} + k_{2}^{-},
&\omega_3 = -4(k_{1}^{+3} + k_{2}^{-3} ),& \Omega_3 =
\sqrt{3}(k_{1}^{+2} -k_{2}^{-2} ) \\k_4 = k_{2}^{+} +
k_{1}^{-},&\omega_4 = -4(k_{2}^{+3} + k_{1}^{-3} ),& \Omega_4 =
\sqrt{3}(k_{2}^{+2} + k_{1}^{-2}),\label{param}
\end{array} \ee
satisfying dispersion relations (\ref{disp}). Substituting these
parameterizations to above solutions we find that \be \alpha_{13}
= 0\,\, \Rightarrow \,\, F_5 = 0,\,\,\,\, \alpha_{23} =
0\,\,\Rightarrow \,\, F_6 = 0 \ee Continuing Hirota's
perturbation with solution $ F_7 = e^{\eta_4} $, where $ \eta _{4}
= k_{4}x + \Omega_{4}y + \omega_{4} t + \eta _{4}^{0}$, we find
that $ F_8 = \alpha_{14}e^{\eta_1 + \eta_4}$,  where \be
\alpha_{14} = -\frac{(k_1 - k_4)(\omega_1 - \omega_4)+ {(k_1 -
k_4)}^4 + {(\Omega_1 - \Omega_4)}^2}{(k_1 +
k_4)(\omega_1+\omega_4)+ {(k_1 + k_4)}^4 + {(\Omega_1 +
\Omega_4)}^2}\ee and after the parameterizations given above
(\ref{param}) it also vanishes \be \alpha_{14} =
0\,\,\Rightarrow\,\, F_8 = 0 \ee The next solution
 $ F_9 =  \alpha_{24}e^{\eta_2 + \eta_4} $,
where \be \alpha_{24} = -\frac{(k_2 - k_4)(\omega_2 - \omega_4)+
{(k_2 - k_4)}^4 + {(\Omega_2 - \Omega_4)}^2}{(k_2 +
k_4)(\omega_2+\omega_4)+ {(k_2 + k_4)}^4 + {(\Omega_2 +
\Omega_4)}^2},\ee  also is zero \be \alpha_{24} = 0
\,\,\Rightarrow\,\, F_9 = 0. \ee Then we have $ F_{10} = 0 $, and
$ F_{11} = \alpha_{34}e^{\eta_3 + \eta_4}$,  where \be
\alpha_{34} = -\frac{(k_3 - k_4)(\omega_3 - \omega_4)+ {(k_3 -
k_4)}^4 + {(\Omega_3 - \Omega_4)}^2}{(k_3 +
k_4)(\omega_3+\omega_4)+ {(k_3 + k_4)}^4 + {(\Omega_3 +
\Omega_4)}^2}.\ee When it is checked for higher order terms we
find that $ F_{12}= F_{13}=...= 0$ . Thus, we have degenerate
four-soliton solution of equations ( \ref{biwadati} ) in the form
 \be F = 1 + e^{\eta _{1}} + e^{\eta
_{2}} + e^{\eta _{3}} + e^{\eta _{4}} + \alpha_{12}e^{\eta
_{1}+\eta _{2}}+ \alpha_{34}e^{\eta _{3}+\eta _{4}}\ee Comparing
this solution with the one in Eq.(\ref{2ssb}) and taking into
account that according parameterizations  (\ref{param}), $\eta_1 +
\eta_2 = \eta_3 + \eta_4 $, we see that they are coincide.
 The above
consideration shows that our two-soliton solution of KP-II
corresponds to the degenerate four soliton solution in the
canonical Hirota form (\ref{biwadati}).
 Moreover, it allows us to find new four virtual soliton
resonance for KPII.

\subsection{Resonance Interaction of Planar Solitons}

Choosing different values of parameters for our two soliton
solution we find resonance character of soliton's interaction. For
the next choice of parameters $k^+_1 = 2, k^-_1 = 1, k^+_2 = 1,
k^-_2 = 0.3$, and vanishing value of the position shift
constants, we obtained two soliton solution moving in the plane
with constant velocity, with creation of the four, so called
virtual solitons (solitons without asymptotic states at infinity).

\begin{figure}[ht]
\begin{center}
\mbox{ \subfigure[Fig.1a]{\includegraphics[height= 6.4cm]{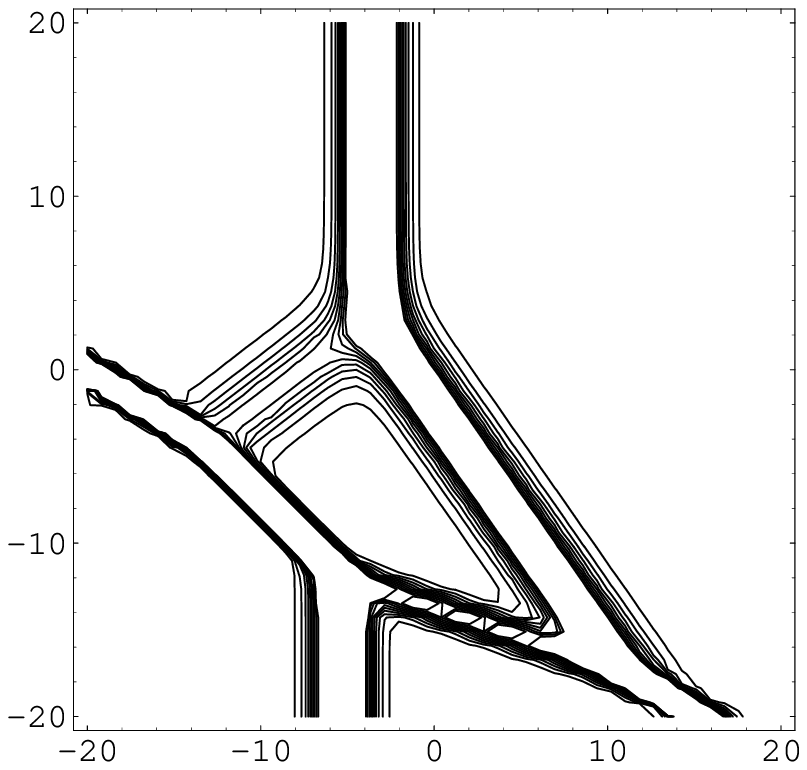}}
\subfigure[Fig.1b]{\includegraphics[height= 6.4cm]{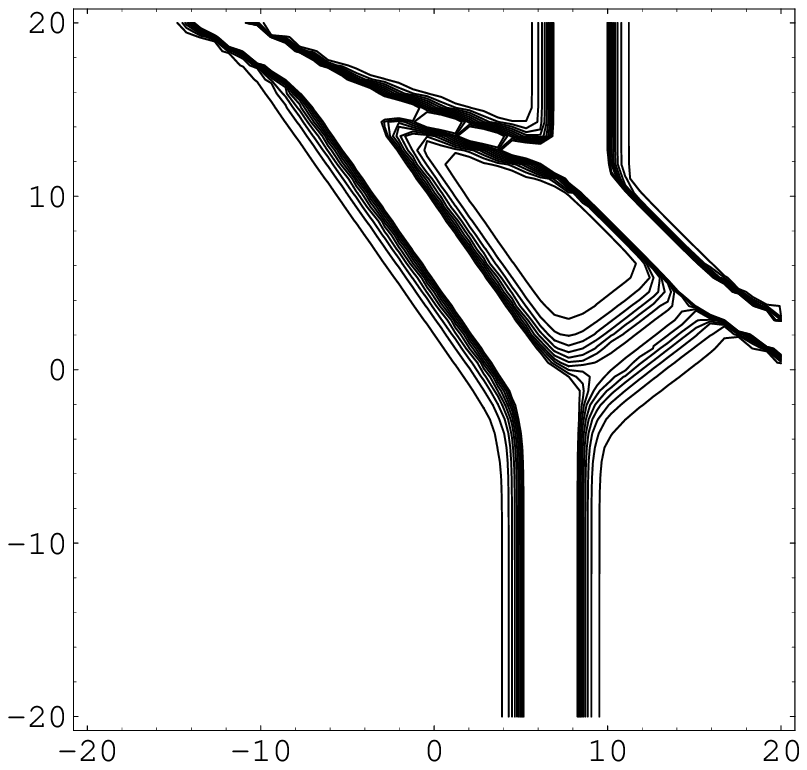}}
 }
\end{center}
\end{figure}

The resonance character of our planar soliton interactions is
related with resonance nature of dissipatons considered in Section
3. It has been reported also in several systems, but the four
virtual soliton resonance does not seem to have been done for
KPII \cite{infeld} prior to our work. In Conference on Nonlinear
Physics, Gallipoli, June-July 2004, we realized that resonance
solitons for KPII very recently have been constructed also by
Biondini and Kodama \cite{BKodama}, \cite{Kodama} using Sato's
theory. Then, the comparision shows that our bilinear constraint
plays the similar role as the Toda lattice in their paper.

\section{Conclusions}

The idea to use couple of equations from the AKNS hierarchy to
generate a solution of KP,  can be applied also to
multidimensional sytems with zero curvature structure as the
Chern-Simons gauge theory. Our three dimensional zero curvature
representation of KP-II gives flat non-Abelian connection for
$SL(2,R)$ and corresponds to a sector of three dimensional
gravity theory. Recently, we have shown that idea similar to the
one presented in this paper can be applied also for Kaup-Newell
hierarchy. In this case, combining the second and the third flows
of dissipative version of derivative NLS (DNLS) we found
resonance soliton dynamics for modified KP-II \cite{PL}.

{\bf Acknoweledgments}
\par\noindent
One of the authors would like to thanks B. Konopelchenko, A.
Pogrebkov, G. Biondini for useful remarks and Y. Kodama for many
valuable discussions clarifying our results. This work was
supported partially by Izmir Institute of Technology, Izmir,
Turkey.

\end{document}